\documentclass[aps,pra,twocolumn,nofootinbib]{revtex4-2}

\usepackage{xcolor}
\usepackage[authormarkup=none, draft]{changes} 

\usepackage{cancel}

\usepackage{graphicx}
\usepackage{subfigure}
\usepackage{amsthm}
\usepackage{amsmath}
\usepackage{verbatim}
\usepackage{dcolumn}
\usepackage{bm}
\usepackage{color}
\usepackage[colorlinks=true,citecolor=blue,linkcolor=blue,urlcolor=blue]{hyperref}%
\usepackage{dsfont}
\usepackage{longtable}
\usepackage{tabularx}
\usepackage{braket}
\usepackage{float,physics}
\usepackage{booktabs}
\usepackage{microtype}
\allowdisplaybreaks
\begin{document}

\title{Quantum Hall correlations in tilted extended Bose-Hubbard chains}
\date{\today}
\author{Hrushikesh Sable}
\email{hsable@vt.edu}
\author{Subrata Das}
\email{subrata@vt.edu}
\author{Vito W. Scarola}
\email{scarola@vt.edu}
\affiliation {Department of Physics, Virginia Tech, Blacksburg, Virginia 24061, USA}

\begin{abstract}  
We demonstrate characteristics of a bosonic fractional quantum Hall (FQH) state in a one-dimensional extended Bose-Hubbard model (eBHM) with a static tilt. In the large tilt limit, quenched kinetic energy leads to emergent dipole moment conservation, enabling mapping to a model generating FQH states. 
Using exact diagonalization, density matrix renormalization group, and an analytical transfer matrix approach, we analyze energy and entanglement properties to reveal FQH correlations. 
Our findings set the stage for the use of quenched kinetics in simple time-reversal invariant eBHMs to explore emergent phenomena. 
\end{abstract}

\maketitle

\noindent
\emph{Introduction---}
Bose-Hubbard models (BHMs) were first constructed to study the quantum liquid phases of helium \cite{MATSUBARA1956a}.  They have since been used to model a variety of systems, including: helium supersolids \cite{MATSUDA1970a,LIU1973a}, disordered superconductors \cite{Fisher1989}, Josephson junction arrays \cite{BRUDER2005}, photonics-based systems (such as photonic crystal cavities and circuit-QED devices \cite{HARTMANN2006,GREENTREE2006,ANGELAKIS2007,HARTMANN2008,CARUSOTTO2009}), and 
 optically trapped ultracold atoms and molecules \cite{jaksch_1998_cold,scarola_2005_quantum,Lewenstein_2007,Bloch2008,chomaz_2022_dipolar}.  Ground states of these models typically follow the Landau paradigm of conventional ordering \cite{Sachdev_2011}.
 
Strategies aimed at enriching the phase diagrams of these models beyond the Landau paradigm frequently employ synthetic external fields, implemented in theoretical frameworks or experimental setups
\cite{Viefers2008,UMUCALILAR2011,NUNNENKAMP2011,dalibard2011,UMUCALILAR2012, grusdt2014, Goldman_2014,monika_2015_artificial, goldman_2016_topological,OZAWA2019,CARUSOTTO2020, Aghtouman2024}. 
Efforts to, for example, construct and study large magnetic field effects on bosons seek to reach the lowest Landau level (LLL) limit where, as Laughlin first pointed out \cite{Laughlin1983}, the flat kinetic energy band leads to interaction-only models captured by idealized parent models \cite{haldane_1983_fractional,trugman1985_exact}. Here, fractional quantum Hall (FQH) states emerge \cite{TSUI1982a,laughlin_1983_anomalous,Jain2007}. The Laughlin states \cite{laughlin_1983_anomalous} defy conventional Landau-paradigm order parameters but are instead defined by a collection of specific features we call FQH correlations: zero energy ground states \cite{haldane_1983_fractional,trugman1985_exact}; robust energy gaps \cite{yoshioka_1983_ground}; gaplesss $U(1)$ Luttinger liquid edges \cite{wen_1991_gapless}; fractionally charged excitations \cite{laughlin_1983_anomalous}; and ground state degeneracies derived from an interplay of many-body translational symmetry \cite{haldane_1985_manyparticlea} and a dipole (or center-of-mass) symmetry \cite{seidel_2005_incompressible}.  The ground state degeneracies connect to topological order \cite{WEN1990a}. For example, a bosonic Laughlin state with a two-fold ground state degeneracy arises in its parent model when the magnetic field is tuned to have two magnetic flux quanta per particle, \emph{i.e.}, at FQH filling  $\nu_{\text{QH}}=1/2$.  

\begin{figure}[t]
    \includegraphics[width=0.9\linewidth]{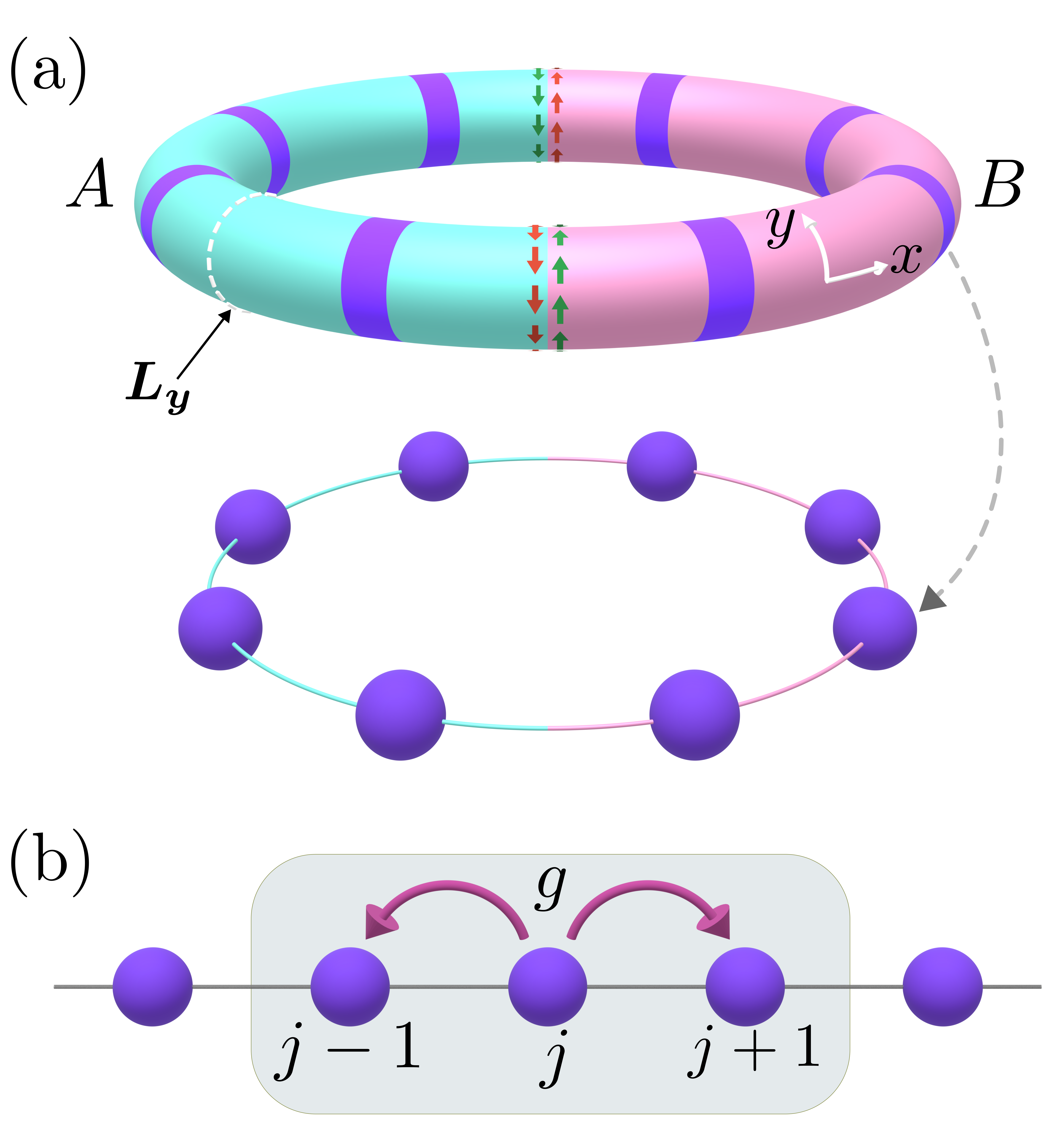}
    \caption{    
    (a) Top: Bosons on a 2D periodic surface in a strong magnetic field perpendicular to the surface (not shown). The circumference is $L_y$.  LLL  single-particle orbitals are drawn as ribbons localized along the $x$-direction. A bipartition separates the system into subsystems A and B used in entanglement calculations.  The arrows at the edges of bipartitions depict FQH edge currents. 
     Bottom: A dipole-conserving lattice model where each site (sphere) is mapped from a corresponding FQH orbital in the top. A similar bipartition divides the lattice. 
    (b) Depiction of the dipole-conserving double hop of two bosons to neighboring sites. }
    \label{fig_schematic}
\end{figure}

The Wannier-Stark effect has recently been explored as a seemingly disparate route to enrich the physics of BHMs \cite{PREISS2015a,yao_2021_manybody,GUO2021b, lake_2022_dipolar,su2023,zechmann_2023_fractonic,ZERBA2025, kim2025}.  Motivated by studies of localization and band flattening \cite{SCHULZ2019,VANNIEUWENBURG2019,KHEMANI2020,scherg_2021_observing,moudgalya_2022_thermalization,will_2024_realization,zechmann_2024_dynamical}, recent work demonstrated that application of a strong static (time-reversal invariant) tilt to the BHM can realize unconventional phases by constraining kinetics and emphasizing interaction induced effects \cite{lake_2023_nonfermi}, in direct analogy to quenched kinetic energy in the LLL \cite{Laughlin1983}.  
The constrained kinetics is due to an emergent dipole-conserving symmetry \cite{lake_2022_dipolar,lake_2023_dipole}, analogous to the symmetry generating FQH correlations. These analogies thus suggest \cite{moudgalya_2022_thermalization} the exciting possibility that mappings [Fig.~\ref{fig_schematic}(a) depicts the mapping] between parent two-dimensional (2D) FQH models and dipole-conserving one-dimensional (1D) lattice models 
\cite{Westerberg1993,bergholtz_2005_halffilleda,seidel_2005_incompressible,seidel_2006_abelian,bergholtz_2006_onedimensional,bergholtz_2008_quantum,lauchli_2010_disentangling,soule_2012_edge,nakamura_2012_exactlya,wang_2013_onedimensional,CRUISE2023a}  might help uncover important but hidden physics in dipole-conserving BHMs.

We connect a  
2D FQH model and a dipole-conserving 1D extended Bose-Hubbard model (eBHM), revealing FQH correlations in the latter. While the conventional Landau-paradigm phases, \emph{e.g.}, density waves (DWs),  have been studied in the 1D eBHM \cite{White2000, CAZALILLA2011}, we demonstrate that the 1D eBHM with a tilt reveals FQH correlations wherein the ground state adiabatically connects to a Laughlin state \cite{rezayi_1994_laughlina,seidel_2005_incompressible}.
We use exact diagonalization (ED) \cite{gaur_2024_exact} and density matrix renormalization group (DMRG)  \cite{white1992,white1993,fishman_itensor_2022,fishman_itensor_2022_03} techniques for numerical simulations, and a matrix product state (MPS) wavefunction \cite{nakamura_2012_exactlya,wang_2013_onedimensional} to analytically compute ground state entanglement properties and make observable predictions toward this emergent quantum phase. Our work thus introduces a surprisingly simple route to realize an FQH correlated state that complements existing approaches in photonic systems \cite{UMUCALILAR2011,NUNNENKAMP2011,UMUCALILAR2012,OZAWA2019}, superconducting qubits \cite{ROUSHAN2017a,kirmani_2022_probing,Kirmani2023}, and ultracold atoms 
\cite{Viefers2008, monika_2015_artificial,goldman_2016_topological,Cooper2019,Leonard_2023_realization}. Our results provide a unifying perspective on a diverse set of topics, connecting topological quantum phases of matter, such as FQH states, with dipole-conserving BHMs.
We summarize by proposing directions to generalize the connection between FQH models and time-reversal invariant BHMs. 

\noindent
\emph{Model---} We consider the repulsive eBHM on a chain with a spatial tilt at lattice filling $\nu_{\text{L}}=1/2$: 
\begin{align}
    \hat{H}_{\rm eBH} &= \sum_{j=1}^{N_s}\frac{U}{2}\hat{n}_j(\hat{n}_j-1)
    + \frac{V}{2} \hat{n}_j\hat{n}_{j+1} +  \Delta j \hat{n}_j \nonumber \\
    &- J( \hat{b}_{j}^{\dagger}\hat{b}^{\vphantom{\dagger}}_{j-1} + {\rm H.c.}), 
    \label{tilt_ebhm}
\end{align}
where $\hat{b}_{j}^{\dagger}$ creates a boson at lattice site $j$, $J$ is the single-particle hopping energy, $U (V)$ is the onsite (nearest-neighbor) interaction energy, $\Delta$ is the tilt strength, and $N_s$ denotes the number of sites.

We expand Eq.~\eqref{tilt_ebhm} in the strong tilt limit, $\Delta \gg J$ and $\Delta \gg U>V$,  using the Schrieffer-Wolff transformation \cite{lake_2023_dipole}.  We set  
${V}/{U} = {2J^2}/{\Delta^2}$ and introduce a gauge transformation on bosonic operators,  
$\hat{a}_j\equiv i^{j\text{mod}2} \hat{b}_j$, to obtain a Hamiltonian valid up to $\mathcal{O}\left[ (J/\Delta)^3, (U/\Delta)^3  \right]$, (proven in Sec.~I, Ref.~\onlinecite{supplemental}):
\begin{align}
   \hat{H}_{g} &= \sum_{j} \bigg[ \hat{n}_j (\hat{n}_j-1) + 2 g (g + 3) \hat{n}_j \hat{n}_{j+1} + g^2 \hat{n}_j \hat{n}_{j+2} \nonumber \\
    +& \left(g\,\hat{a}_{j}^{\dagger}\hat{a}_{j+1}^2\hat{a}_{j+2}^{\dagger}  
    - g\,^2 \, \hat{a}_{j-1}^{\dagger}\hat{a}_{j}^{\vphantom{\dagger}}\hat{a}_{j+1}^{\vphantom{\dagger}}\hat{a}_{j+2}^{\dagger} + {\rm H.c.}\right) \bigg],
  \label{short_range_laughlin_mapped}
\end{align}
where the gauge transformation gives a positive dipole hopping term [the second to last term as depicted in Fig.~\ref{fig_schematic}(b)] with strength: $g \equiv 2 J^2/( \Delta^2 - 2J^2)$. 
Importantly, $\hat{H}_g$ commutes with the dipole-moment operator $\hat{P} = \sum_j j \hat{n}_j$ (mod $N_s$). We therefore consider Eq.~\eqref{short_range_laughlin_mapped} as projected into individual dipole moment sectors, thus allowing omission of the constant tilt term and use of periodic boundary conditions.

Equation~\eqref{short_range_laughlin_mapped} has another  important symmetry: $[\hat{H}_g, \hat{T}]=0$, where $\hat{T}=\prod_{j=1}^{N_s} \hat{T}_j$ is a many-body translational operator,  and $\hat{T}_j$ translates a particle at site $j$ by one site to the right. $\hat{T}$ is of the same form as the symmetry \cite{haldane_1985_manyparticlea} underlying topological order in Laughlin states \cite{WEN1990a} and establishes a FQH-like algebra: $\hat{U} \hat{T} = {\rm exp}(2\pi i \nu_{\text{L}})\hat{T}\hat{U}$,  where $\hat{U}\equiv{\rm exp}(2\pi i \hat{P}/N_s )$.  The phase factor arising from the analogous FQH algebra is, for comparison, ${\rm exp}(2\pi i \nu_{\text{QH}})$ \cite{seidel_2005_incompressible}. 
Because $\hat{T}$ and $\hat{U}$ both commute with $\hat{H}_g$ but do not commute with each other, for $\nu_{\text{L}}=1/2$, both $\ket{\psi_g}$ and $\hat{T} \ket{\psi_g}$ have different eigenvalues of $\hat{U}$, leading to a two-fold degenerate ground state sector \cite{oshikawa_2000_commensurability,seidel_2005_incompressible}, analogous to $\nu_{\text{QH}}=1/2$ FQH models. 
The existence of these symmetries in $\hat{H}_g$ establishes underlying conditions for FQH correlations, as we argue next.

We compare the low energy properties of Eq.~\eqref{short_range_laughlin_mapped} to a 2D FQH parent model \cite{haldane_1983_fractional,soule_2012_edge} of bosons with a repulsive delta function interaction on a thin cylinder of circumference $L_y$ and periodic boundaries at $\nu_{\rm QH}=1/2$:
\begin{align}
    \hat{H}_{\rm QH} \,{=} \sum_{j=1}^{N_s} \sum_{k \ge |m|} 
    e^{-\frac{2\pi^2(k^2 + m^2)}{L_y^2}}  \hat{B}_{j+m}^{\dagger}\hat{B}_{j+k}^{\dagger}\hat{B}_{j+m+k} \hat{B}_{j},
    \label{untruncated_laughlin_hamil}
\end{align}
where $\hat{B}_{j}^{\dagger}$ creates a boson in LLL orbital $j$ [see Fig.~\ref{fig_schematic}(a), and Sec.~II of Ref.~\onlinecite{supplemental} for review].

The matrix elements are exponentially suppressed in $1/L_y^2$ and thus allow expansion by a small parameter.  To see how the small parameter arises, note that a single particle LLL basis state can be written as a product of a plane wave around the circumference ($y$-direction) and a Gaussian along the length ($x$-direction): $e^{i (2\pi k/L_y)\, y}\phi_m(x)$, where $k$ is an integer and $\phi_k(x)\sim{\rm \exp}[{-(x+2\pi k/L_y)^2/2]}$.  The strength of the FQH Hamiltonian matrix elements is then determined by the overlaps between nearest-neighbor basis states along $x$: $\vert \langle \phi_k(x)\vert \phi_{k\pm 1}(x)\rangle\vert^2\sim {\rm exp}(-4 \pi^2/L_y^2)$, which yields a small parameter for low $L_y$.  For large $L_y$, the ground states of $\hat{H}_{\rm QH}$ become the bosonic Laughlin state \cite{bergholtz_2006_onedimensional}.  
 
Remarkably, the first four terms of Eq.~\eqref{short_range_laughlin_mapped} are nearly the same as the four lowest-order terms of Eq.~\eqref{untruncated_laughlin_hamil} (Sec.~II of Ref.~\onlinecite{supplemental}). The coefficient of the nearest-neighbor interaction term in  Eq.~\eqref{short_range_laughlin_mapped} differs from the FQH model 
and the last term in Eq.~\eqref{short_range_laughlin_mapped} is qualitatively distinct; otherwise they are the same. This shows that $\hat{H}_g$ has the form of a short-ranged FQH-like model. To equate the matrix elements of both models, we make the assignment $g\rightarrow 2{\rm exp}(-4 \pi^2/l^2)$, where $l$ is an artificial length parameter, in Eq.~\eqref{short_range_laughlin_mapped}.  By inserting $l$, we quantitatively connect a \emph{length} scale in a 2D FQH model (circumference, $L_y$) to \emph{internal parameters} in $\hat{H}_g$ (the ratio of energy scales captured by $l$).  Conversely, the internal parameters of $\hat{H}_g$ can be used to study the length scaling in FQH models. 
As both $J$ and $\Delta$ are highly controllable in synthetic quantum systems such as cold atoms in optical lattices, tuning $g$ is feasible.
 
\noindent
\emph{Spectral Structure---} 
We compare properties of eigenstates of Eqs.~\eqref{short_range_laughlin_mapped} and \eqref{untruncated_laughlin_hamil} in finite-size systems. 
\begin{figure}[tbp]
    \centering
    \includegraphics[width=0.49\textwidth]{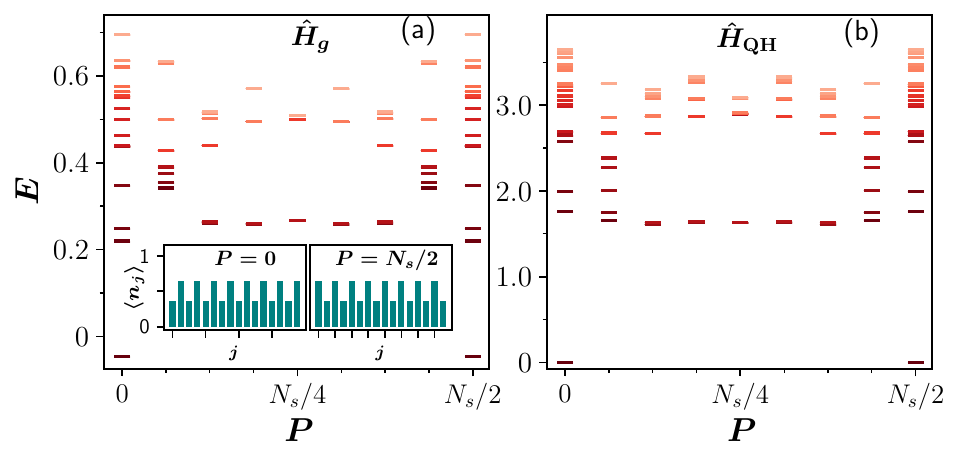}
    \caption{
    (a) Low-lying energy eigenvalues of $\hat{H}_g$ versus dipole moment eigenvalue, $P$, at $g=0.8$ for $N_s=16$. Here we see the ground state degeneracy between $P$ sectors and the energy gap. The insets depict the density profile for the ground states in $P=0$ and $P=N_s/2$ sectors. (b) The same but for $\hat{H}_{\text{QH}}$ at $L_y = 6.5$.
    } 
    \label{fig:density_ebhm}
\end{figure}
As expected from Eq.~\eqref{untruncated_laughlin_hamil}, we first checked that Eq.~\eqref{short_range_laughlin_mapped} has two zero-energy ground states for low $g$, using ED and DMRG for up to $32$ particles. At high $g$, while the degeneracy is still two, there are small deviations from zero energy due to the last term in Eq.~\eqref{short_range_laughlin_mapped}. This confirms a two-fold degenerate ground state and energy gap (Sec.~I of Ref.~\cite{supplemental} presents additional charge gap data).  These findings are consistent with generic requirements \cite{oshikawa_2000_commensurability} in lattice models at $\nu_L=1/2$.  

To see the impact of differences between Eqs.~\eqref{short_range_laughlin_mapped} and \eqref{untruncated_laughlin_hamil}, we simulate moderate to large $g$ values.
Figure~\ref{fig:density_ebhm} shows example data comparing energy spectra of Eqs.~\eqref{short_range_laughlin_mapped} and \eqref{untruncated_laughlin_hamil} for moderately high $g$, $g=0.8$. As noted earlier, the ground state is marginally lowered below zero, by $\sim 10^{-2}$, implying that the correction in energy is an order of magnitude smaller than $g$.
There are other changes to the excitation spectra.  For instance, comparing panels (a) and (b) of Fig.~\ref{fig:density_ebhm}, we see that the $P=1$ excitation sector is higher for Eq.~\eqref{short_range_laughlin_mapped}. Nonetheless, we have checked that the spectra remain qualitatively similar as we tune $g$.

\begin{figure}[btp]
    \centering
    \includegraphics[width=0.44\textwidth]{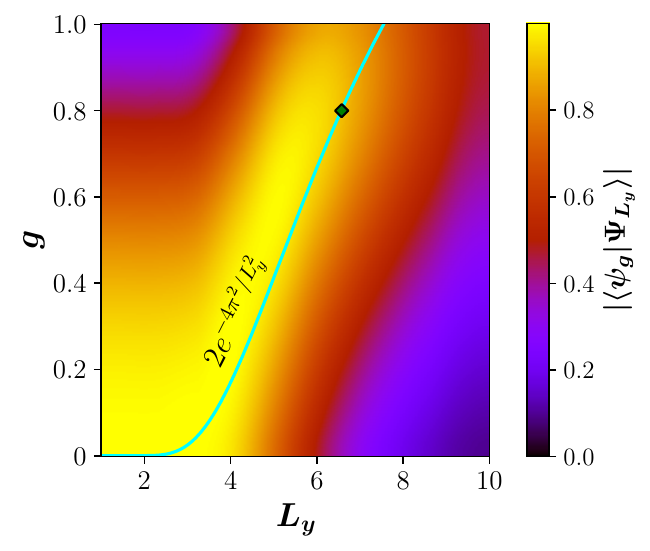}
    \caption{Overlap between the ground states of $\hat{H}_g$ and $\hat{H}_{\text{QH}}$ as we vary dipole hopping strength in $\hat{H}_g$ against the torus circumference in the FQH parent model, for $N_s = 16$.  The solid line denotes the parameter choice where the lowest-order terms in each model match. The diamond denotes an example parameter $g=0.8$ choice that yields an overlap of $90\%$. } 
    \label{fig:olap}
\end{figure}

\noindent
\emph{Ground State as an MPS---} 
We now focus on ground state properties. Figure~\ref{fig:olap} shows the overlap, 
$|\Bra{\psi_g} \Psi_{L_y}\rangle|$, between the ground state of Eq.~\eqref{untruncated_laughlin_hamil}, $\ket{\Psi_{L_y}}$, and the ground state of Eq.~\eqref{short_range_laughlin_mapped}, $\ket{\psi_g}$, computed using ED.  The solid line depicts the path obtained by equating $l$ and $L_y$ to best approximate the matrix elements of both models.  Along this line, we find significant overlaps, nearly $90\%$ and higher, for $L_y \lesssim 7$. The overlap starts to decrease at large $g$ because higher-order terms in Eq.~\eqref{untruncated_laughlin_hamil} and the last term in Eq.~\eqref{short_range_laughlin_mapped} differ.  We note that the gauge prescription used to make the dipole hopping term positive is crucial for a non-zero overlap. This implies that $\ket{\psi_g}$ features FQH correlations up to this gauge transformation.

The ground state of Eq.~\eqref{short_range_laughlin_mapped} can, in the absence of the $g^2$ hopping term, be written as an MPS \cite{nakamura_2012_exactlya}: 
\begin{equation}
   \ket{\psi_{g}}_{\text{MPS}} = \prod_j \left[1 - \frac{g}{\sqrt{2} } \hat{a}_{j-1}(\hat{a}_{j}^{\dagger})^2\,\hat{a}_{j+1}\right]\ket{101010\ldots}.
   \label{nakamura_gs_main}
\end{equation}
This wavefunction is equivalent to a bosonic Laughlin state \cite{wang_2013_onedimensional} up to linear order in $g$.  $\ket{\psi_{g}}_{\text{MPS}}$ allows analytic computation of correlation functions with the transfer matrix method.  

Correlation functions based on the density reveal a DW phase featuring quantum correlations. $\ket{\psi_{g}}_{\text{MPS}}$ and our numerical results for $\ket{\psi_{g}}$ on finite-size systems show oscillating density order, as in Fig.~\ref{fig:density_ebhm}.  We also find that density-density fluctuations, $\langle n_j n_{j'}\rangle-\langle n_j\rangle \langle n_{j'}\rangle$ decay exponentially with $\vert j-j'\vert$, consistent with expectations of a gapped 1D quantum phase \cite{HASTINGS2006gap} (Sec.~III of Ref.~\onlinecite{supplemental} provides numerical data). However, this DW has key differences from ordinary DWs.  Most prominently, the eigenstates of Eq.~\eqref{short_range_laughlin_mapped} have zero dipole moment fluctuations, $\langle \hat{P}^2\rangle-\langle \hat{P}\rangle^2=0$ arising from FQH symmetries, whereas conventional DWs studied in typical eBHMs [\emph{e.g.}, Eq.~\eqref{tilt_ebhm} with $\Delta=0$ \cite{White2000}] have $\langle \hat{P}^2\rangle-\langle \hat{P}\rangle^2\neq0$.  Sec.~IV of Ref.~\onlinecite{supplemental} shows an example comparison.  Dipole moment fluctuations, therefore, offer an observable that distinguishes dipole-conserving ground states from conventional DWs.

\noindent
\emph{Bipartite Entanglement and Density Fluctuations---} 
We now characterize the entanglement properties of $\hat{H}_g$ ground states.  The substitution $g\rightarrow 2{\rm exp}(-4 \pi^2/l^2)$ allows us to also examine the scaling of entanglement properties with $l$ in direct comparison with conventional studies of entanglement that typically probe system-size (length scale) dependence \cite{eisert_2010_colloquium}.  We start with the entanglement spectrum (ES).  

The ES $\{\xi_n\}$ is defined in terms of the pure state Schmidt decomposition of the ground state:
$\sum_{n}e^{-\xi_n/2} \ket{\psi_n^A}\otimes \ket{\psi_n^B}$.
The states $\ket{\psi_n^A}\,\, (\ket{\psi_n^B})$ form an orthonormal basis for subsystem $A \,\, (B)$ due to the bipartition, depicted in Fig.~\ref{fig_schematic}(a).
The ES can also be labeled with a quantum number defined by the dipole-moment eigenvalue for partition A, $P_{\text{A}} (\text{mod} N_s/2)$ \cite{li_2008_entanglement,lauchli_2010_disentangling}. 

\begin{figure}[btp]
    \centering
    \includegraphics[width=0.49\textwidth]{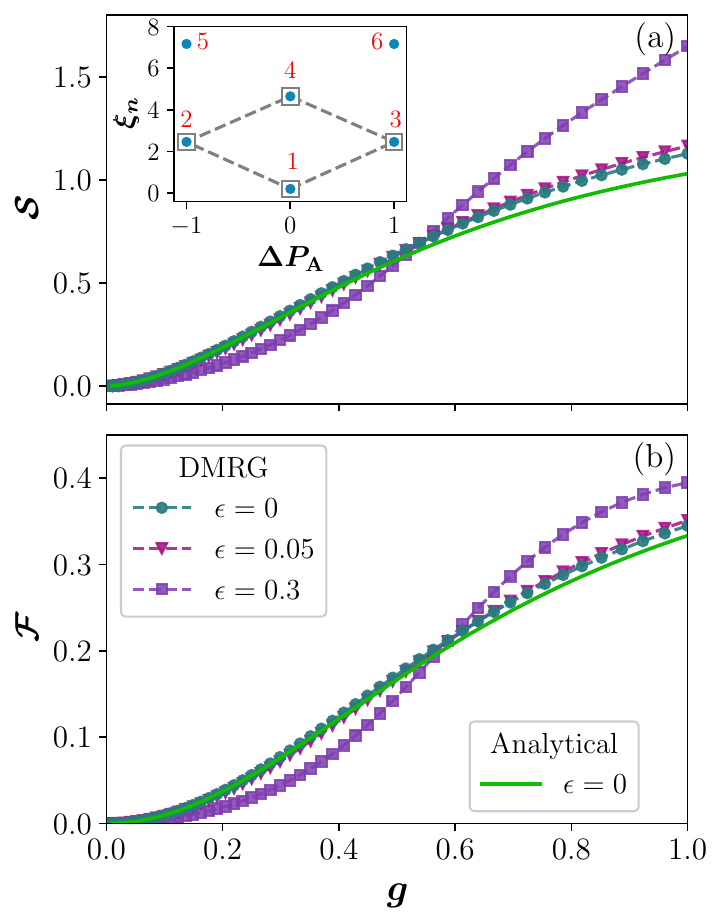}
    \caption{(a) Entanglement entropy versus $g$, showing both DMRG data (symbols) for the ground state of $\ket{\psi_{g}}$ and analytical result from $\ket{\psi_{g}}_{\text{MPS}}$ (solid line) for $N_s = 64$. 
    The inset depicts the ES, $\xi_n$, versus dipole moment difference for the ground state of $\hat{H}_{\text{QH}}$ (empty squares) and the ground state of $\hat{H}_{g}$ (filled circles) for $g = 0.5$, and adjacent numbers represent $n$.  
    The dashed lines are a guide. (b) The same, but for the bipartite number fluctuations.  
    }
    \label{fig:entropy_fluc}
\end{figure}

The inset of Fig.~\ref{fig:entropy_fluc}(a) plots the ES against $\Delta P_{\text{A}}$, the deviation in dipole moment from the``vacuum" state defined as $\ket{101010\ldots}$.  
The two largest eigenvalues arise from the last (correction) term in $\hat{H}_g$.  Otherwise, the remaining four lowest levels form a diamond where the ES of $\hat{H}_g$ and the FQH model match.  To understand the four-level diamond ES structure, we start with $\ket{\psi_1^A}$, the state adiabatically connected to the vacuum state with eigenvalue $\xi_1$ near zero.  Removing or adding an edge particle to $\ket{\psi_1^A}$ creates two degenerate dispersing modes, with energies $\xi_2$ and $\xi_3$ that arise at $\Delta P_{\text{A}}=\pm 1$ [these map to the linearly dispersing edge modes in the FQH model depicted as edge currents in Fig.~\ref{fig_schematic}(a)].  The fourth state, with energy $\xi_4$, can be thought of as a combination of these dispersing modes so that the energy is roughly the sum of the energies of both dispersing modes and the $\Delta P_{\text{A}}$ combine to cancel, leaving  $\Delta P_{\text{A}}=0$.  The diamond structure is therefore consistent with the conformal tower structure of edge Tomonaga-Luttinger liquids seen for FQH models
\cite{lauchli_2010_disentangling,wang_2013_onedimensional}.   Note that the edge of $\hat{H}_g$ is just a single lattice site but is nonetheless characterized by FQH-like edge current ES with $l$.

We use $\ket{\psi_{g}}_{\text{MPS}}$ to derive the ES. Sec.~V of Ref.~\onlinecite{supplemental} reports the full formulas.  Expanding about $g = 0$ up to $\mathcal{O}(g^3)$ yields: $\xi_1\sim g^2$, $\xi_2=\xi_3\sim {\rm ln}(2)-2{\rm ln}(g)+2g^2$, and $\xi_4\sim 2{\rm ln}(2)-4{\rm ln}(g)+3g^2$.  Taking $g\rightarrow 2{\rm exp}(-4 \pi^2/l^2)$ shows that the ES diamond has a height $\xi_4-\xi_1= {\rm ln}(4)+16\pi^2l^{-2}+\mathcal{O}[{\rm exp}(-1/l^2)]$, where the $l^{-2}$ term shows the leading order impact of entanglement as the eBHM hopping increases (or, similarly, as the 
FQH torus diameter is enlarged).  The quadratic scaling of $\xi_1$ with $g$  manifests in other entanglement quantities as well. 

We also compute the entanglement entropy:  $\mathcal{S} = \sum_n \xi_n e^{-\xi_n}$ \cite{eisert_2010_colloquium}.
Figure~\ref{fig:entropy_fluc}(a) compares $\mathcal{S}$ for $\ket{\psi_{g}}$, and the analytic result obtained from $\ket{\psi_{g}}_{\text{MPS}}$ (see Sec.~V of Ref.~\cite{supplemental} for the full functional form of $\mathcal{S}$), and  
we see excellent agreement. 
Since $\ket{\psi_{g}}_{\text{MPS}}$ is equivalent to a bosonic Laughlin state at linear order in $g$, the agreement of the numerical data of $\hat{H}_g$ with analytics reiterates the FQH correlations in $\hat{H}_g$. Some deviations occur at large $g$  due to the last term in Eq.~\eqref{short_range_laughlin_mapped}.
Furthermore, we have checked that $\mathcal{S}$ for Eq.~\ref{untruncated_laughlin_hamil} is identical to within $3 \%$ for $g\lesssim 0.8$. 

We extract the asymptotic scalings of the entanglement entropy using 
$\ket{\psi_{g}}_{\text{MPS}}$. We find that the $g^2$ and ${\rm ln}(g)$ scaling in $\xi_n$ appear such that: $\mathcal{S} (g) = g^2 \left[1 + {\rm ln}(2) - {\rm ln}(g^2)\right] + \mathcal{O}(g^4)$.  For $l\rightarrow 0$, we find a vanishing entanglement entropy: $\mathcal{S} \rightarrow {\rm exp}(-8\pi^2/l^2)[1+{\rm ln}(2)+8\pi^2/l^2)]$, consistent with the vanishing entanglement on a thinning FQH torus model. We find that the large and small $l$ limits of $\mathcal{S}$ saturate to constants, which is also consistent with a theorem establishing area law bounds on the entanglement entropy for gapped short-range 1D spin models \cite{Hastings_2007_area}.  The saturation contrasts with known 2D area law scalings of entanglement entropy, $\sim c l$, where $c$ is a constant \cite{eisert_2010_colloquium}. 

The scaling of the entanglement spectrum and entropy can be connected with observables \cite{song_2012_bipartite,cristian_2025_entanglement}.  We compute the bipartite number fluctuations of subsystem $A$ defined as: $\mathcal{F} =  \langle (\hat{N}_A - \langle \hat{N}_A\rangle )^2 \rangle$, 
with the expectation value taken with respect to the ground state  \cite{song_2012_bipartite}.  We find $\mathcal{F}(g) = g^2 + \mathcal{O}(g^4)$, which results from the $g^2$ scaling found in $\xi_1$.

To check the robustness of these predictions, we introduce $\epsilon$: $V/U = 2J^2/\Delta^2 + \epsilon$. Since $\hat{H}_g$ is derived for $\epsilon=0$, a non-zero $\epsilon$ perturbs all the terms of $\hat{H}_g$ (proven in Sec.~VI in Ref.~\onlinecite{supplemental}).
Fig.~\ref{fig:entropy_fluc} shows that $\epsilon$ does not change the predictions for $\mathcal{S}$ and $\mathcal{F}$ as long as it is well below the energy gap.  The scaling of the observable $\mathcal{F}$ with $g$ can therefore be used to verify the predicted thin torus FQH scaling of $\mathcal{S}$ with $g$ in the eBHM.

\noindent
\emph{Outlook---}
We demonstrate FQH correlations hidden in strongly tilted 1D eBHMs.
Future work shall examine other possible emergent properties imposed by tilt 
including searches for a non-trivial Zak phase \cite{Zak_1989,Niu_2010}, calculations done at different fillings to explore the concept of vortex attachment \cite{Jain2007,scarola2014}, and a possible analog to the bosonic Moore-Read state \cite{Moore1991a,MA2024324} at $\nu_{\text{L}}=1$.  It would also be interesting to establish bounds on the minimal number of longer-range interaction terms needed in the 1D eBHM to observe a topological parameter scaling: $\sim c\,l+\gamma$, where $\gamma$ is 
topological entanglement entropy, and has value $\log(d)$ where $d$ denotes the ground-state degeneracy \cite{levin_2006_detecting,kitaev_2006_topological, lauchli_2010_entanglement}. In addition, higher-dimensional eBHMs might reveal similar FQH correlations by connecting 4D FQH models \cite{FROHLICH2000,ZHANG2001} to eBHMs with time reversal invariant fields.

\noindent
\emph{Acknowledgments-} H.S. thanks Lo\"\i c Herviou for useful discussions. We acknowledge support from  AFOSR (FA9550-19-1-0272, FA2386-21-1-4081, and FA9550-23-1-0034) and ARO (W911NF2210247).  

\emph{Data availability} The data that support the findings of this article are openly available at \cite{sable_2026_data}.

\bibliography{references}

\end{document}


\title{Supplementary material for ``Quantum Hall correlations in tilted extended Bose-Hubbard chains"}
\author{Hrushikesh Sable}
\author{Subrata Das}
\author{Vito W. Scarola}
\affiliation {Department of Physics, Virginia Tech, Blacksburg, Virginia 24061, USA}
\maketitle

\section{Derivation of $\hat{H}_g$ from $\hat{H}_{\rm eBH}$}
In this section, we shall outline the derivation of $\hat{H}_g$ starting from a one-dimensional (1D) extended Bose-Hubbard model (eBHM) with a tilt, given by Eq.~(1) in the main text.  In the limit $\Delta \gg J$ and $\Delta \gg U > U$, using the Schrieffer-Wolff transformation, $\hat{H}_{\rm eBH}$ is expanded in powers of $J/\Delta$ and $U/\Delta$. As discussed in the main text, $\Delta$ denotes the tilt strength, while $J$, $U$, and $V$ represent the single-particle hopping, the onsite interaction, and the nearest-neighbor (NN) interaction energy, respectively. Expanding upto $\mathcal{O}\left[ (J/\Delta)^3, (U/\Delta)^3  \right]$, one obtains the following model \cite{lake_2023_dipole}:

\begin{eqnarray}
    \hat{H}  &=& -\sum_{j}\Big[\frac{J^2(U - V)}{\Delta^2} \hat{b}_{j}^{\dagger}\hat{b}_{j+1}^2 \hat{b}_{j+2}^{\dagger} + \frac{J^2 V}{\Delta^2} \hat{b}_{j-1}^{\dagger}\hat{b}_{j}\hat{b}_{j+1}\hat{b}_{j+2}^{\dagger} + {\rm H.c.}\Big] + \sum_j \Big[\Delta j \hat{n}_j -\frac{U}{2}\hat{n}_j \Big]\nonumber \\
    && + \sum_{j}\Bigg[\Big\{\frac{U}{2} - \frac{2J^2(U - V/2)}{\Delta^2}\Big\}\hat{n}_j^2 + \Big\{V + \frac{4J^2(U - V)}{\Delta^2}\Big\}\hat{n}_j  \hat{n}_{j+1} + \frac{J^2 V}{\Delta^2} \hat{n}_j  \hat{n}_{j+2} \Bigg].
    \label{hamil_DPeBHM}
\end{eqnarray}
This shows that in the strong tilt limit, we obtain a dipole preserving Hamiltonian, Eq.~\eqref{hamil_DPeBHM}, and so we have $[\hat{H},\hat{P}] = 0$, where $\hat{P} = \sum_j j \hat{n}_j$ (mod $N_s$)
is the dipole moment operator. Setting $U$ as the energy scale of $\hat{H}$, and introducing the gauge-transformation on bosonic operators: $\hat{a}_j\equiv i^{j\text{mod}2} \hat{b}_j$, we obtain
\begin{eqnarray}
    \hat{H}  &=& \sum_{j}\Bigg[\frac{J^2}{\Delta^2}\left(1 - \frac{V}{U}\right) \hat{a}_{j}^{\dagger}\hat{a}_{j+1}^2 \hat{a}_{j+2}^{\dagger} -\frac{J^2 V}{\Delta^2 U} \hat{a}_{j-1}^{\dagger}\hat{a}_{j}\hat{a}_{j+1}\hat{a}_{j+2}^{\dagger} + {\rm H.c.}\Bigg] \nonumber \\ &+&  \sum_{j}\Bigg[\Bigg\{\frac{1}{2} - \frac{2J^2}{\Delta^2}\left(1 - \frac{V}{2U}\right)\Bigg\}\hat{n}_j (\hat{n}_j - 1) 
    +  \left\{\frac{V}{U} + \frac{4J^2}{\Delta^2} \Big(1-\frac{V}{U}\Big)\right\}\hat{n}_j\hat{n}_{j+1} + \frac{J^2 V}{\Delta^2 U} \hat{n}_j \hat{n}_{j+2}\Bigg]
    \label{hamil_DPeBHM_reparam1}
\end{eqnarray}
where we have:
(i) dropped the $ \sum_j \Delta j \hat{n}_j$ term since the Hamiltonian commutes with dipole moment $\hat{P}$, and 
(ii) expressed the $\hat{n}_j^2$ term as  $\hat{n}_j (\hat{n}_j -1)$ and have ignored the $\hat{n}_j$ term (which is an effective chemical potential) as it does not modify the correlations.  Note that Eq.~\eqref{hamil_DPeBHM_reparam1} is valid for any $J/U$. The first term involves three sites: $j, j+1, j+2$, and has a correlated form such that the hopping from $j+1$ to $j$ is accompanied by the hopping from $j+1$ to $j+2$, preserving the dipole moment $\hat{P}$. We shall refer to this hopping as dipole hopping. 

\begin{figure}[htbp]
    \centering
    \includegraphics[width=0.49\textwidth]{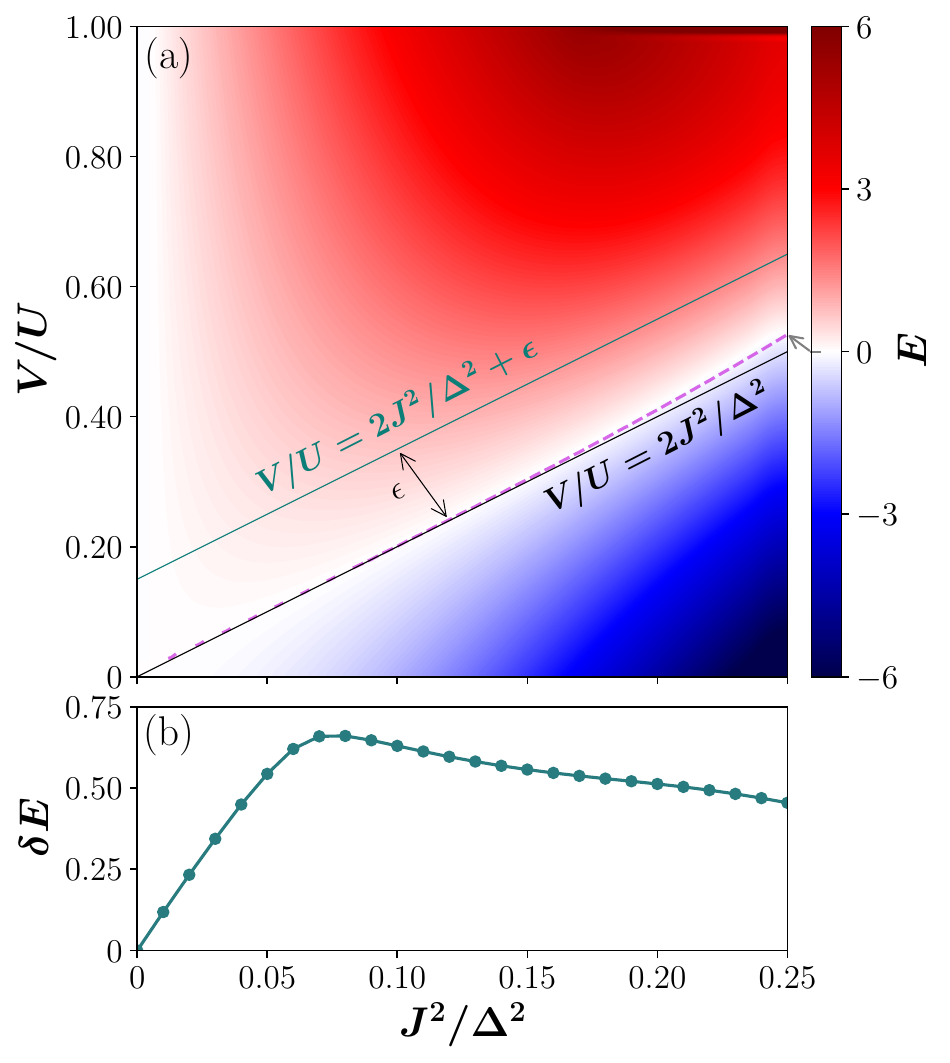}
    \caption{(a) Ground state energy of Eq.~\eqref{hamil_DPeBHM_reparam1} as a function of $J^2/\Delta^2$ and $V/U$, for $N_s = 64$ sites. The zero energy contour is shown as a dashed red line, and the $V/U=2J^2/\Delta^2$ line is denoted in solid black. The parameter $\epsilon$ indicates perturbation from the $V/U = 2J^2/\Delta^2$ limit, see Sec.~\ref{hg_perturb}.
    (b) Charge excitation gap as a function of $J^2/\Delta^2$ along the $V/U = 2J^2/\Delta^2$ line, indicating a gapped phase. } 
    \label{fig:energy}
\end{figure}

As discussed in the main text, we are interested in signatures of $1/2$ filling bosonic fractional quantum Hall (FQH)  states in Eq.~\eqref{hamil_DPeBHM_reparam1}. Accordingly, we use the fact that Laughlin states are exact ground states of the Haldane pseudopotentials \cite{haldane_1983_fractional, trugman1985_exact}.  See Eq.~\eqref{untruncated_laughlin_hamil} below for reference. In particular, for the bosonic $\nu_{\text{QH}}=1/2$ Laughlin state, the criterion is that the dipole hopping coefficient squared equals the product of coefficients of $\hat{n}_j (\hat{n}_j - 1)$ and the $\hat{n}_j  \hat{n}_{j+2}$ term. We impose this criterion on the coefficients in Eq.~\eqref{hamil_DPeBHM_reparam1} and obtain: 
$$V/U = 2J^2/\Delta^2.$$ 
In this limit, Eq.~\eqref{hamil_DPeBHM_reparam1} becomes $\hat{H}_g$ [Eq.~(2) in the main text], with the definition: 
\begin{align}
g =  2 J^2 / (\Delta^2 - 2J^2), 
\end{align}
as stated in the main text.

To verify that $V/U=2J^2/\Delta^2$ indeed leads to FQH correlations, we compute the characteristics of the ground states, using both exact diagonalization (ED) \cite{gaur_2024_exact} and the density-matrix renormalization group (DMRG) technique \cite{fishman_itensor_2022}. Fig.~\ref{fig:energy}(a) plots the ground state energy $E$ of Eq.~\eqref{hamil_DPeBHM_reparam1} for lattice filling $\nu_{\rm L} = 1/2$. We notice a distinct region of zero energy, forming along the $V/U=2J^2/\Delta^2$ line.  Note that a zero-energy ground state is a characteristic of a $1/2$ bosonic FQH state. 
Upon comparing $\hat{H}_g$ with the truncated FQH model - Eq.~\eqref{trunc_model} below, we find that the two are qualitatively similar. The key differences lie in the NN interaction and the $g^2$ hopping term. Although the NN interaction has a different coefficient in both models, it does not affect the ground state manifold. In contrast, the $g^2$ hopping term in $\hat{H}_g$ has no counterpart in Eq.~\eqref{trunc_model}, which accounts for the deviations from $E = 0$ observed at large values of $J^2/\Delta^2$.

In addition to $E\approx 0$ ground state, our numerics also show a non-zero charge gap.  We define the charge gap as: $$\delta E = E(N+1,N_s) - 2 E(N,N_s) + E(N-1,N_s),$$ where $E(N,N_s)$, is ground state energy of $\hat{H}_g$ for $N$ bosons on $N_s$ sites.  Fig.~\ref{fig:energy}(b) plots the gap. This is consistent with the expectation that FQH states have gapped charge excitations.

\section{Review of fractional quantum Hall models}
\label{fqh_review}
The FQH models describe interacting particles in a two-dimensional (2D) $x$-$y$ plane subjected to a perpendicular magnetic field $\mathbf{B} = B \hat{z}$. In the Landau gauge, the vector potential is $\mathbf{A} = B\, x \hat{y}$. On a cylinder with transverse (circumferential) length $L_x$ ($L_y$), the single-particle orbitals in the lowest Landau level (LLL) are \cite{bergholtz_2005_halffilleda,seidel_2005_incompressible,Jain2007}:
\begin{equation}
    \phi_k (\mathbf{x}) \propto e^{i (2\pi k/L_y)\, y} e^{-1/2(x + 2\pi k/L_y)^2}.  
    \label{single_part_orb}
\end{equation}
For a long thin cylinder ($L_x \rightarrow \infty$) subjected to periodic boundary conditions, the essential properties become those of a torus geometry.  Here $2\pi k/L_y$ is the single-particle momentum along the transverse direction and $k$ is an integer. 
The number of such available single particle orbitals is $N_s = L_y L_x/2\pi $. These orbitals are quasi-periodic and centered at $x_k = -2\pi k/L_y$, as shown by distinct ribbons in Fig.~1(a) in the main text. This implies that the momentum along the $y$ direction describes the central position along $x$. Using this single-particle basis, the matrix elements for the interaction in the second-quantized form can be constructed \cite{Jain2007,soule_2012_edge,wang_2013_onedimensional}.

The Laughlin state of bosons at $\nu_{\text{QH}} = 1/2$ filling is the exact zero energy ground state of the two-particle interaction \cite{haldane_1983_fractional, trugman1985_exact}:
\begin{equation}
    \mathcal{V}(\mathbf{x}) = \delta^2(\mathbf{x}),
    \label{bosonic_pseudopot}
\end{equation}
in the LLL. When the single particle basis states given by Eq.~\eqref{single_part_orb} are used to obtain a second-quantized form for the interaction potential in Eq.~\eqref{bosonic_pseudopot}, one obtains the following model \cite{nakamura_2012_exactlya, wang_2013_onedimensional}:
\begin{eqnarray}
    \hat{H}_{\rm QH} &=& \sum_{j=1}^{N_s} \sum_{k \ge |m|} {\rm exp}\left[-\frac{2\pi^2(k^2 + m^2)}{L_y^2}\right]\hat{B}_{j+m}^{\dagger}\hat{B}_{j+k}^{\dagger}\hat{B}_{j+m+k} \hat{B}_{j},
    \label{untruncated_laughlin_hamil}
\end{eqnarray}
where $\hat{B}_{j} (\hat{B}_{j}^{\dagger})$ annihilates (creates) a boson at LLL orbital $j$, and the assumption $L_x \rightarrow \infty$  simplifies the interaction matrix elements \cite{wang_2013_onedimensional}. 
Thus, in the Landau gauge, $\hat{H}_{\rm QH}$ is an effective 1D model in the LLL, and is long-ranged in nature \cite{seidel_2005_incompressible,bergholtz_2005_halffilleda}. This Hamiltonian describes the processes in which two bosons with separation $k+m$ move $m$ orbitals in opposite directions, and the strength of that process is given by the exponential term in the parenthesis. Note that in Eq.~\eqref{untruncated_laughlin_hamil}, there is a symmetry in swapping the annihilation operators ($\hat{B}_{j+m+k} \hat{B}_{j} \leftrightarrow \hat{B}_{j}\hat{B}_{j+m+k}$) if the orbitals are distinct, and similarly for the creation operators. One therefore needs to consider numerical prefactors in front of the terms based on combinatorics.

Since the interaction matrix elements are exponentially suppressed in $1/L_y^2$, we can retain dominant terms in Eq.~\eqref{untruncated_laughlin_hamil}. In particular, by restricting $k + |m| \leq 2$, we obtain a truncated FQH model that has four lowest order terms:
\begin{eqnarray}
    \hat{H}_{\rm TQH} &=& \sum_{j=1}^{N_s}\Big[\hat{N}_j(\hat{N}_j-1) + 4 \widetilde{V}\,\hat{N}_j  \hat{N}_{j+1} +  4 \widetilde{V}^4\, \hat{N}_j  \hat{N}_{j+2} + 2 \widetilde{V}^2\, (\hat{B}_{j-1}^{\dagger}\hat{B}_j^{2}\hat{B}_{j+1}^{\dagger}  + {\rm H.c.}) \Big],
    \label{trunc_model}
\end{eqnarray}
where $\widetilde{V} \equiv e^{-(2\pi^2/L_y^2)}$ and $\hat{N}_j \equiv \hat{B}_{j}^{\dagger}\hat{B}_j$.

To identify the counterpart of $L_y$, the circumference length that governs the interactions in the $\hat{H}_{\rm QH}$ (and $\hat{H}_{\rm TQH}$), in the Hamiltonian $\hat{H}_{g}$, we present a side-by-side comparison of the coefficients of the dipole-hopping terms in both models, see Table.~\ref{table_param}. In analogy with $L_y$, we introduce an artificial length parameter $l$ that is an invertible function of $g$, which brings the two models on the same footing. It is by this parametrization that we notice a region of highest overlap in Fig.~(3) in the main text.

\renewcommand{\arraystretch}{1.25} 
\begin{table}[htbp]
\caption{Table indicating the coefficients of the dipole hopping and the parametrization in terms of the length variable.}
\label{table_param}
\begin{ruledtabular}
\begin{tabular}{ccc}
& $\hat{H}_{\rm QH}$ &  $\hat{H}_{g}$ \\
\hline \\
Length Parameter & $L_y$ & $l$ \\
Hamiltonian Parameter & $2 \widetilde{V}^2$ & $g$   \\
Hamiltonian Parameter vs. Length Parameter & $2 \widetilde{V}^2 = 2{\rm exp}(-4\pi^2/L_y^2)$ & $g = 2{\rm exp}(-4\pi^2/l^2)$ \\
\end{tabular}
\end{ruledtabular}
\end{table}

\section{Additional data for density-density correlations}
In this subsection, we provide additional data for density-density correlations.  As seen in Fig.~2 in the main text, the ground state of $\hat{H}_g$ has a density wave (DW) pattern and has quantum correlations for $g\ne 0$. To investigate the nature of these quantum correlations, we compute the density-density correlation function: 
$C(\vec{r}_i,\vec{r}_j) = \langle \hat{n}_i \hat{n}_j \rangle - \langle \hat{n}_i \rangle \langle \hat{n}_j \rangle$.

\begin{figure}[hbtp]
    \centering
    \includegraphics[width=0.5\textwidth]{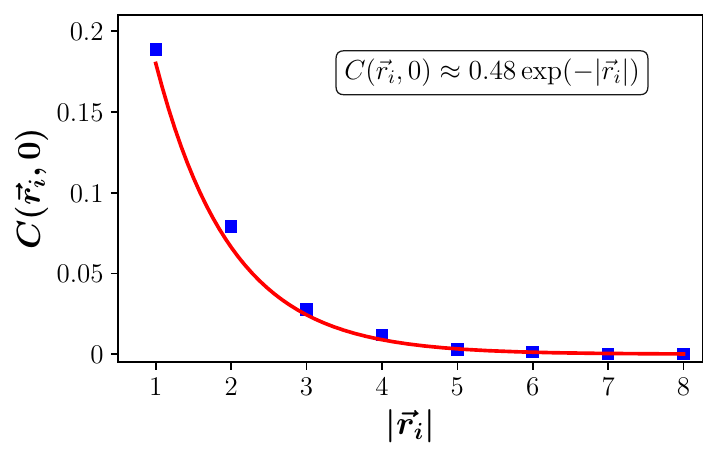}
    \caption{Plot of $C(\vec{r}_i,0)$ as a function of $|\vec{r}_i|$ for the ground state of $\hat{H}_g$ for $g = 0.6$ and $N_s = 16$. The blue squares (red line) are the numerical ED (analytical) data.} 
    \label{fig:den_den_corr}
\end{figure}

Figure~\ref{fig:den_den_corr} illustrates the exponential decay of $C(\vec{r}_i,0)$ with $|\vec{r}_i|$. We use both - ED and analytical Matrix Product State (MPS) formalism (see Sec.~\ref{mps_analytics} below), to obtain the results, and notice an excellent agreement between the two. The exponential decay of $C(\vec{r}_i,0)$ is consistent with the expectation for a 1D gapped quantum phase \cite{HASTINGS2006gap}, and suggests short-ranged quantum correlations on top of the DW order in the quantum phase. 

\section{Numerical data for dipole moment fluctuations}
\begin{figure}[htbp]
    \centering
    \includegraphics[width=0.5\textwidth]{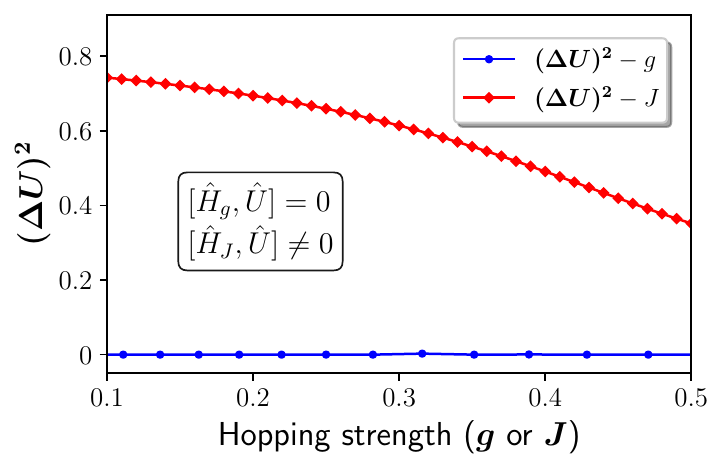}
    \caption{The dipole moment fluctuations  $(\Delta U)^2$ for the two models, $\hat{H}_g $ (shown by blue circles) and  $\hat{H}_J$ (shown by red diamonds), as a function of the respective hopping strengths, computed with ED for $N_s=16$. } 
    \label{fig:com_fluc}
\end{figure}

Figure~2 in the main text and Fig.~\ref{fig:den_den_corr} suggest that the quantum phase we study has a DW pattern with short-ranged quantum fluctuations imprinted onto it. In this section, we present data that distinguishes this correlated DW from an analogous correlated DW state that one would obtain considering a conventional eBHM, which has the usual single particle hopping. In particular, we consider a conventional eBHM:
\begin{align*}
\hat{H}_{J} = \sum_{j} \hat{n}_j (\hat{n}_j - 1) + 2 J (J + 3) \, \hat{n}_j  \hat{n}_{j+1} + J^2 \, \hat{n}_j  \hat{n}_{j+2} + J\left(\, \hat{b}_{j}^{\dagger}\hat{b}_{j+1}  + {\rm H.c.}\right), 
\end{align*}
where $\hat{b}_j$ ($\hat{b}^{\dagger}_j$) is the bosonic annihilation (creation) operator for site $j$. $\hat{H}_{J}$ is the same as $\hat{H}_g$, except we replace the dipole hopping terms with the usual single-particle hopping term, to make a consistent comparison. We fix the lattice filling, $\nu_{\rm L} = 1/2$ for this calculation as well.  

We consider the exponentiated dipole moment operator $\hat{U}\equiv{\rm exp}(2\pi i \hat{P}/N_s )$ and demonstrate the dipole moment fluctuations: $(\Delta U)^2 = \langle \hat{U}^2 \rangle - \langle \hat{U} \rangle^2 $, for both $\hat{H}_g$ and $\hat{H}_J$ in Fig.~\ref{fig:com_fluc}. Since $\hat{H}_g$ is of dipole-conserving form, $(\Delta U)^2 = 0$, whereas $(\Delta U)^2 \neq 0$ for $\hat{H}_J$, due to the mixing of different dipole-moment sectors induced by single particle hopping. Thus $(\Delta U)^2$ distinguishes between the DWs of $\hat{H}_g$ and of $\hat{H}_J$.

\section{Analytical derivations using Matrix Product state wavefunction}
\label{mps_analytics}
The ground state of $\hat{H}_{\rm TQH}$ is an MPS \cite{nakamura_2012_exactlya}:
\begin{equation}
   \ket{\psi_{g}}_{\text{MPS}} = \prod_j \left[1 - \frac{g}{\sqrt{2} } \hat{a}_{j-1}(\hat{a}_{j}^{\dagger})^2\,\hat{a}_{j+1}\right]\ket{101010\ldots}.
   \label{nakamura_gs_supp}
\end{equation}
(Note that we have used the equality correspondence between bosonic operators $\hat{a}_j$ and $\hat{B}_j$).
It was shown that the densities, correlation functions, and the entanglement entropy can be obtained analytically for the short-ranged FQH model using the transfer matrix method \cite{wang_2013_onedimensional}. Since $\hat{H}_g$ is essentially the same as $\hat{H}_{\rm TQH}$ upto a $g^2$ hopping correction, we can use the analytical expressions derived using the MPS representation and compare them with the numerically computed observables using the ground state of $\hat{H}_g$. The state in Eq.~\eqref{nakamura_gs_supp} can be mapped to a spin-1 model using the following unit cell mapping: $\ket{10} \rightarrow \ket{\bar{o}}$, $\ket{00} \rightarrow \ket{-}$ and $\ket{02} \rightarrow \ket{+}$. With this spin-1 notation, Eq.~\eqref{nakamura_gs_supp} can be written as:
\begin{equation}
 \ket{\psi_g}_{\rm MPS} = \frac{1}{\sqrt{\mathcal{N}}}{\rm tr}   [M_0 M_1 M_2 \ldots M_{N-1}] 
  \label{nakamura_gs_spin}
\end{equation}
where $N$ denotes number of bosons and $\mathcal{N}$ is the normalization. The matrix $M_Q$ is 
\begin{equation}
  M_Q = 
  \begin{bmatrix}
  \ket{\bar{o}} & \ket{+} \\
  -\frac{g}{\sqrt{2}} \ket{-}& 0
  \end{bmatrix}
\end{equation}
for all $Q$, where $Q$ denotes the unit cell index ranging from $0$ to $N-1$, and the normalization is then given as $\mathcal{N} = {\rm tr}[G^N]$ where $G = \overline{M}_Q \otimes M_Q$ is the $4\times 4$ transfer matrix. The two non-zero eigenvalues of this transfer matrix are $\{(1 \pm \sqrt{1 + 2g^2})/2\}$.

\subsection{Density}
\begin{figure}[htbp]
    \centering
    \includegraphics[width=0.5\textwidth]{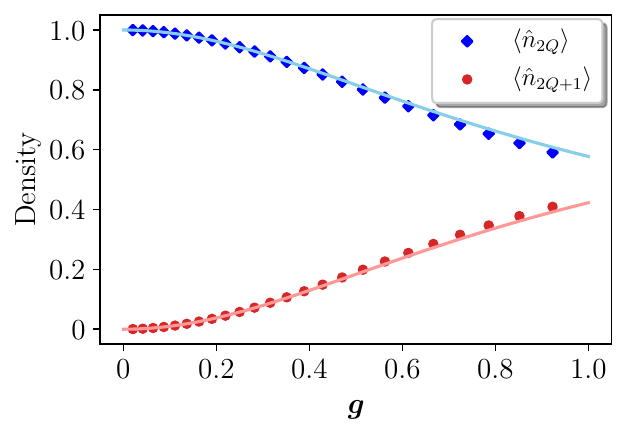}
    \caption{Ground state densities of $\hat{H}_{g}$ (solid circles) computed numerically for a system of $N_s = 16$ sites, and those given in Eq.~\eqref{density_analy} (solid lines). The blue diamonds (red circles) denote the density at even (odd) sites.} 
    \label{fig:nakamura_analy_density}
\end{figure}

Using Eq.~\eqref{nakamura_gs_spin}, the density at odd and even sites can be obtained analytically. We first express the density operator as a matrix product operator:
\begin{equation}
    \hat{n}_{2Q + q} = (q+1)\delta(S^z_Q, q).
\end{equation}
where $q = 0$ ($q=1$) corresponds to even (odd) site within a unit cell, and $S_Q^{z}$ means the $z$ component of spin-1 $Q$th unit cell. In this notation, a lattice site $j$ can be expressed as: $j = 2Q + q$. Using the transfer matrix $G$ and its eigenvalues, the densities at even and odd sites, in the large $N \rightarrow \infty$ limit, are: 
\begin{equation}
\langle \hat{n}_{2Q + q} \rangle =
\begin{cases}
  \dfrac{1}{\sqrt{1 + 2g^2}} & \text{for } q = 0, \\[8pt]
  1 - \dfrac{1}{\sqrt{1 + 2g^2}} & \text{for } q = 1
\end{cases}
\label{density_analy}
\end{equation}
We compare these analytical expressions for the density with the numerical data of the $\hat{H}_{g}$, and as seen in Fig.~\ref{fig:nakamura_analy_density}, we find an excellent agreement. Since truncated FQH models, such as Eq.~\ref{trunc_model}, are exactly solvable using MPS formalism, modeling the density of $\hat{H}_g$ with MPS indirectly confirms the FQH correlations in $\hat{H}_g$. At large $g$, there are small deviations since the $g^2$ hopping term starts to play a role.

\subsection{Entanglement spectrum and entanglement entropy}

In this subsection, we outline the steps to compute the entanglement entropy $\mathcal{S}$ for the wavefunction in Eq.~\eqref{nakamura_gs_spin}. As discussed in the main text, we divide the system into two equal halves to compute $\mathcal{S}$. The Schmidt decomposition that illustrates this bipartition is: 
\begin{equation}
    \ket{\Psi} = \sum_{n}e^{-\xi_n/2} \ket{\psi_n^A}\otimes \ket{\psi_n^B},
\end{equation}
for a pure state $\ket{\Psi}$.  The states $\ket{\psi_n^A}\,\, (\ket{\psi_n^B})$ form an orthonormal basis for subsystem $A \,\, (B)$ on account of bipartition.

The set $\{\xi_n\}$ is referred to as the entanglement spectrum (ES). $\mathcal{S}$ is defined as the von-Neumann entropy of the reduced subsystem:
\begin{equation}
    \mathcal{S} = -{\rm tr} (\rho_A {\rm ln} \rho_A)
    \label{entropy_def}
\end{equation}
where $\rho = \vert \Psi \rangle \langle \Psi\vert$ is the total density matrix, and $\rho_A = {\rm tr}_B \rho$ is the reduced density matrix.
In terms of the ES, the entropy $\mathcal{S}$ can be expressed as: $\mathcal{S} = \sum_n \xi_n e^{-\xi_n}$.

Since the ground state is an MPS, the reduced density matrices after the bipartition also have a product form however, now only over the subsystem and with different normalization. Using these, the ES can be obtained. 
In the infinite-size limit, we obtain: 
\begin{eqnarray}
  \xi_2 &=& \xi_3 = {\rm ln}\left(4 + \frac{2}{g^2}\right) \nonumber \\
  \xi_1 &=& \ln\!\left(4+\frac{2}{g^2}\right)
+ \ln\!\left(\frac{\sqrt{2g^2+1}-1}{\sqrt{2g^2+1}+1}\right)   \nonumber \\
  \xi_4 &=& \ln\!\left(4+\frac{2}{g^2}\right)
- \ln\!\left(\frac{\sqrt{2g^2+1}-1}{\sqrt{2g^2+1}+1}\right).  
\label{ent_spec_nakamura_s}
\end{eqnarray}
Using these expressions, the entanglement entropy can be expressed as
\begin{widetext}
\begin{align}
\mathcal{S}(g) = -\frac{1}{(1 + 2g^2)\left(1 + \sqrt{1 + 2g^2}\right)^2} \Bigg[
  & 2g^2\left(1 + g^2 + \sqrt{1 + 2g^2}\right)
  \ln\left(\frac{g^2}{2 + 4g^2}\right)
  + g^4 \ln\left\{ \frac{g^4}{(1 + 2g^2)\left(1 + \sqrt{1 + 2g^2}\right)^2} \right\}
  \nonumber \\
  & + \left\{ g^4 + 2\left(1 + \sqrt{1 + 2g^2}\right) + 2g^2\left(2 + \sqrt{1 + 2g^2}\right) \right\}
  \ln\left\{ \frac{\left(1 + \sqrt{1 + 2g^2}\right)^2}{4 + 8g^2} \right\}
\Bigg].
\label{entropy_nakamura_s}
\end{align}
\end{widetext}
As we notice in Fig.~4(a) in the main text, there is an excellent agreement with the numerically obtained values.

In the main text, we have plotted the ES against $\Delta P_A$, which is the change in the dipole moment in subsystem $A$, measured with respect to the vacuum state defined as $\ket{101010\ldots}$. The subsystem dipole moment $P_A$ is a conserved quantity that enables us to label the ES with it. To understand the ES, we list the following scenarios which alter the configuration at the edges (due to the dipole hopping), label the corresponding $P_A$ next to it, and the ES level. To simplify the analysis, we first ignore the $g^2$ dipole hopping term in $\hat{H}_g$ (or equivalently, consider $\hat{H}_{\rm TQH}$).

\begin{table}[H]
\centering
\caption{The lowest $4$ different configurations at edges and corresponding ES levels, only due to dipole hopping.}
\begin{tabular}{c|c|c c}
\rule{0pt}{3ex}$B$ & $A$ & $B$ \\
\rule{0pt}{3ex} \ldots $1\,0$ & $1\,0\,1\,0\,1\,0$ & $1\,0\,1\,0\ldots$ &  $\quad \quad P_A = 3$ $\quad \quad \xi_1$  \\
\rule{0pt}{3ex} \ldots $0\,2$ & $0\,0\,1\,0\,1\,0$ & $1\,0\,1\,0\ldots$ &  $\quad \quad P_A = 2$ $\quad \quad \xi_2$  \\
\rule{0pt}{3ex} \ldots $1\,0$ & $1\,0\,1\,0\,0\,2$ & $0\,0\,1\,0\ldots$ &  $\quad \quad P_A = 4$ $\quad \quad \xi_3$ \\
\rule{0pt}{3ex} \ldots $0\,2$ & $0\,0\,1\,0\,0\,2$ & $0\,0\,1\,0\ldots$ &  $\quad \quad P_A = 3$ $\quad \quad \xi_4$ \\
\end{tabular}
\label{table_es_nakamura}
\end{table}

Table~\ref{table_es_nakamura} lists the $4$ ES for the ground state of $\hat{H}_{\rm TQH}$, reported in Eq.~\eqref{ent_spec_nakamura_s}. This explains two degenerate levels: $\xi_2$ and $\xi_3$, at $\Delta P_A = \pm 1$. Furthermore, the levels $\xi_1$ and $\xi_4$ correspond to $\Delta P_A = 0$ level, and $\xi_4$ is obtained by combining two dispersing modes at $\Delta P_A = +1$ and $\Delta P_A = -1$. This explains the diamond structure in Fig.~4 in the main text. 

The next dominant ES are listed in Table~\ref{table_higher_es}.  These are raised due to the additional $g^2$ term present in $\hat{H}_{g}$, which can be understood by operating the $g^2$ hopping onto the states written in Table.\ref{table_es_nakamura}. This explains Fig.~4 in the main text, where we see $2$ additional levels above the diamond.

\begin{table}[H]
\caption{The additional configurations at edges due to $g^2$ hopping and corresponding ES levels.}
\centering
\begin{tabular}{c|c|c c}
\rule{0pt}{3ex}$B$ & $A$ & $B$ \\
\rule{0pt}{3ex} \ldots $1\,0$ & $1\,0\,1\,0\,0\,1$ & $1\,1\,0\,0\ldots$ &  $\quad \quad P_A = 4$ $\quad \quad \xi_5$  \\
\rule{0pt}{3ex} \ldots $0\,1$ & $1\,1\,0\,0\,0\,1$ & $1\,1\,0\,0\ldots$ &  $\quad \quad P_A = 2$ $\quad \quad \xi_6$ 
\label{table_higher_es}
\end{tabular}
\end{table}

\section{Derivation of model with perturbed parameters}
\label{hg_perturb}
We check the robustness of the ground state physics of $\hat{H}_g$ by introducing a perturbation parameter $\epsilon$: $V/U = 2J^2/\Delta^2 + \epsilon $. The perturbed $\hat{H}_{g}(\epsilon)$ is, 
\begin{eqnarray}
    \hat{H}_{g}(\epsilon) &=& \sum_{j}\Bigg[  \Big\{1 + \epsilon g\left(1 + g\right)\Big\} \hat{n}_j (\hat{n}_j - 1) + \Big\{2 g (g +3) + 2 \epsilon \, (1 - g^2 )\Big\} \hat{n}_j  \hat{n}_{j+1} + \Big\{g^2 + \epsilon\,g(1+g)\Big\} \hat{n}_j  \hat{n}_{j+2} \Bigg]\nonumber \\
    &+&\sum_{j}\Big[\big\{g - \epsilon\, g\left(1+g\right)\big\} \hat{b}_{j}^{\dagger}\hat{b}_{j+1}^2\hat{b}_{j+2}^{\dagger} -  \big\{g^2 + \epsilon\,g(1+g)\big\} \hat{b}_{j-1}^{\dagger}\hat{b}_{j}\hat{b}_{j+1}\hat{b}_{j+2}^{\dagger} + {\rm H.c.}\Bigg]
\end{eqnarray}
This shows that this perturbation is different from one used in usual perturbation theory, as it modifies all the terms of the Hamiltonian. We consider different values of $\epsilon$ and notice that the observable properties remain constant (Fig.~4 in the main text) as long as $\epsilon$ is below the gap. 

\section{Computational methods}
We briefly discuss the computational methods that we have employed in our work. In particular, we have performed ED and DMRG studies for the Hamiltonians under study. ED is employed using the Lanczos algorithm for diagonalization for system sizes ranging up to $16$ sites, with a maximum of $4$ bosons occupying each site. The dipole-conserving symmetry of the problem enables us to reduce the matrix dimension. 

The DMRG algorithm is implemented in the ITensor C++ library \cite{fishman_itensor_2022}. We set the maximum bond dimension to be $\mathcal{D}=1600$, and we truncate the eigenvalues below a cut-off of $10^{-12}$. This truncation significantly reduces the effective bond dimension required. For instance, the simulation of the short-ranged interacting Hamiltonian for a system of $64$ sites requires a maximum bond dimension of $200$ for a converged MPS.  The convergence criterion is set to $10^{-10}$ for the energy difference between two consecutive sweeps. The system size is varied from $N_s=8$ to $N_s=64$ sites, and we use periodic boundary conditions in Hamiltonian construction.

\bibliography{references}